\documentclass[prl,twocolumn,showpacs,preprintnumbers,amsmath,amssymb,superscriptaddress]{revtex4}
\usepackage{graphicx}
\usepackage{dcolumn}
\usepackage{bm}

\begin{document}

\title{Dirac fermions at the $H$ point of graphite: Magneto-transmission studies}
\author{M. Orlita}
\email{orlita@karlov.mff.cuni.cz} \affiliation{Grenoble High Magnetic Field
Laboratory, CNRS, BP 166, F-38042 Grenoble Cedex 09, France}
\affiliation{Institute of Physics, Charles University, Ke Karlovu 5, CZ-121~16
Praha 2, Czech Republic} \affiliation{Institute of Physics, v.v.i., ASCR,
Cukrovarnick\'{a} 10, CZ-162 53 Praha 6, Czech Republic}
\author{C. Faugeras}
\affiliation{Grenoble High Magnetic Field Laboratory, CNRS, BP 166, F-38042
Grenoble Cedex 09, France}
\author{G. Martinez}
\affiliation{Grenoble High Magnetic Field Laboratory, CNRS, BP 166, F-38042
Grenoble Cedex 09, France}
\author{D. K. Maude}
\affiliation{Grenoble High Magnetic Field Laboratory, CNRS, BP 166, F-38042
Grenoble Cedex 09, France}
\author{M. L. Sadowski}
\affiliation{Grenoble High Magnetic Field Laboratory, CNRS, BP 166, F-38042
Grenoble Cedex 09, France}
\author{M. Potemski}
\affiliation{Grenoble High Magnetic Field Laboratory, CNRS, BP 166, F-38042
Grenoble Cedex 09, France}
\date{\today}

\begin{abstract}
We report on far infrared (FIR) magneto-transmission measurements on a
thin graphite sample prepared by exfoliation of highly oriented
pyrolytic graphite (HOPG). In magnetic field, absorption lines
exhibiting a blue-shift proportional to $\sqrt{B}$ are observed.
This is a fingerprint for massless Dirac holes at the $H$ point in
bulk graphite. The Fermi velocity is found to be
$\tilde{c}=(1.02\pm0.02)\times10^6$~m/s and the pseudogap $|\Delta|$ at the $H$
point is estimated to be below 10~meV. Although the holes behave
to a first approximation as a strictly 2D gas of Dirac fermions,
the full 3D band structure has to be taken into account to
explain all the observed spectral features.
\end{abstract}


\maketitle

The fabrication of graphene, a 2D lattice of carbon atoms with a
honeycomb symmetry, and the subsequent discovery of Dirac
fermions~\cite{NovoselovScience04,BergerJPCB04,NovoselovNature05,ZhangNature05}, has
lead to renewed interest in the physical properties of bulk
graphite. Compared to graphene, graphite represents a system of a
higher complexity which is still not fully understood despite
fifty years of intensive research. The currently accepted tight
binding (TB) model, formulated for graphite by Slonzewski, Weiss
and McClure (SWM)~\cite{SlonczewskiPR58,McClurePR60}, implies
seven TB parameters $\gamma_{0},\ldots\gamma_{5}, \Delta$ and
predicts the presence of particles with parabolic and linear
dispersions at the $K$ and $H$ points of the Brillouin zone
respectively. Whereas the appearance of massive electrons at the
$K$ point has been reported in numerous experiments, see e.g.
\cite{GaltPR56,SchroederPRL70,LiPRB06}, there is little direct
evidence for Dirac fermions (holes) at the $H$ point.

Evidence for $\sqrt{B}$-dependent features in magneto-reflectance
spectra, typical for Dirac particles, has been reported by
Toy~\emph{et al.}~\cite{ToyPRB76}. Recently, quantum oscillations
in the magnetoresistance of bulk graphite
~\cite{LukyanchukPRL04,LukyanchukPRL06}, indicated the presence of
strictly 2D gases of both Dirac holes and massive electrons.
Nevertheless, this interpretation remains
controversial~\cite{MikitikPRB06,BernevigPRL07} and raises some
doubts concerning the accuracy of the SWM model. Further evidence
of Dirac fermions close to the $H$ point was obtained from angle
resolved photoemission spectroscopy (ARPES)
~\cite{ZhouNatPhys06,GruneisPRL08}. However, the ARPES measurements
gave differing Fermi velocities, information on the band structure
only below the Fermi level, and are characterized by a rather low
accuracy. Dirac holes have recently been observed in scanning
tunneling spectroscopy (STS)~\cite{LiNatPhys07}, indicating again
the pronounced 2D behavior of particles in HOPG. However, these
investigations are principally limited to the surface of the
sample and it is not understood why these effects have not been
observed in equivalent experiments performed
earlier~\cite{MatsuiPRL05}.

\begin{figure}
\scalebox{0.8}{\includegraphics*[175pt,211pt][414pt,411pt]{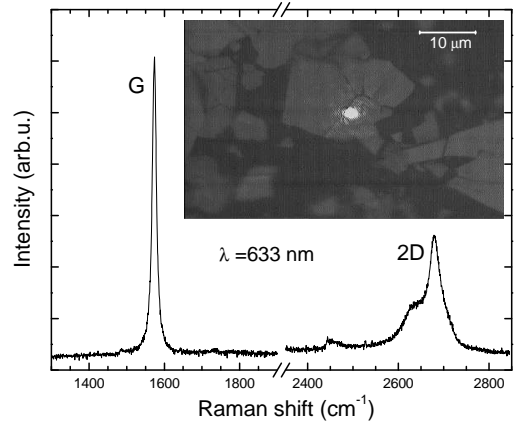}}
\caption{\label{Raman} A Raman spectrum of the studied sample
showing the Raman bands $G$ and $2D$ typical of bulk
graphite~\cite{FerrariPRL06}. The bright laser spot in the middle
of the microscope image in the inset shows the place, where this
spectrum was collected.}
\end{figure}

In this Letter, we use FIR magneto-transmission experiments to
probe the nature of the holes at the $H$ point of HOPG. We address
the currently controversial issue of whether the holes in HOPG can
be described using a simplified model assuming a strictly 2D gas
of Dirac fermions as proposed in
Refs.~\cite{LukyanchukPRL04,LukyanchukPRL06,LiNatPhys07}, or
whether a full 3D band structure needs to be employed. We show
that the appealing 2D model is in the simplest approach
applicable, but simultaneously, we demonstrate the clear limits of
this approximation. In this work, we focus on transitions, the
energies of which scale as $\sqrt{B}$. However, in a different
spectral region (low energies and high magnetic fields) we clearly
observe features linear in $B$, which in accordance to previous
reports~\cite{LiPRB06} arise from electronic transitions in the
vicinity of the $K$ point.

\begin{figure}
\scalebox{1.15}[1.15]{\includegraphics*[92pt,520pt][307pt,720pt]{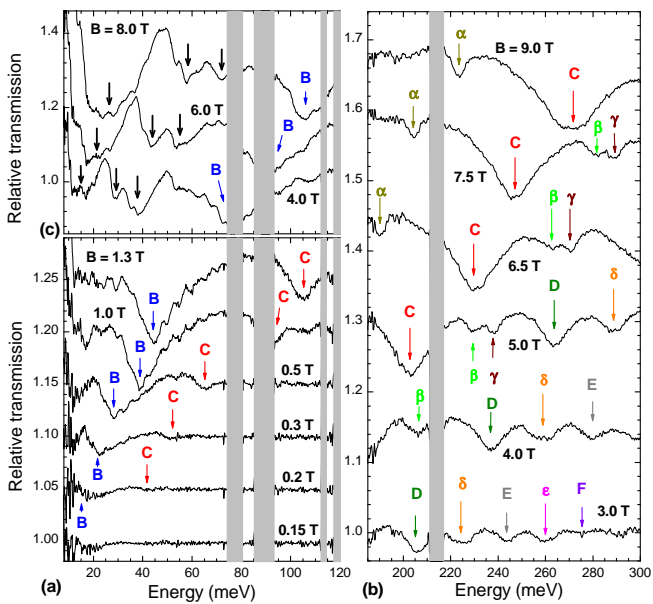}}
\caption{\label{SPKT} (color online) Transmission spectra at
selected magnetic fields in the low (a),(c) and high (b) energy
windows. The lines denoted by individual letters exhibit a
$\sqrt{B}$-scaled blue shift and are related to the $H$ point. The
low-energy features in (c), nearly linear in $B$, originate from
the $K$ point. The black arrows in (c) denote the most developed transmission minima
whose positions were plotted into the inset of Fig.~\ref{LL}.
For clarity, successive spectra in (a),(b) and (c) are shifted
vertically by 0.05, 0.15 and 0.15, respectively.}
\end{figure}

A thin sample for the transmission measurements was prepared by
exfoliation of HOPG. A confocal microscope image of part of the
sample, i.e. of the tape with stacked graphite layers, is shown in
the inset of Fig.~\ref{Raman}. Scanning across the sample in the micro-Raman
experiment, see Fig.~\ref{Raman}, we only detected the signal typical of bulk
graphite~\cite{FerrariPRL06}. This signal corresponds to the light areas in the picture,
whereas the other dark areas (tape) exhibited no graphite or graphene signal. The
layers of the bulk graphite of various thickness cover roughly
50\% of the tape surface. The transmission experiment was
performed on a macroscopic round-shaped sample having 5~mm in
diameter. The average thickness of the graphite layers, $\simeq
100$~nm, was roughly estimated from the transmissivity of the
sample in the visible range.

\begin{figure}
\scalebox{0.85}{\includegraphics*[16pt,15pt][294pt,233pt]{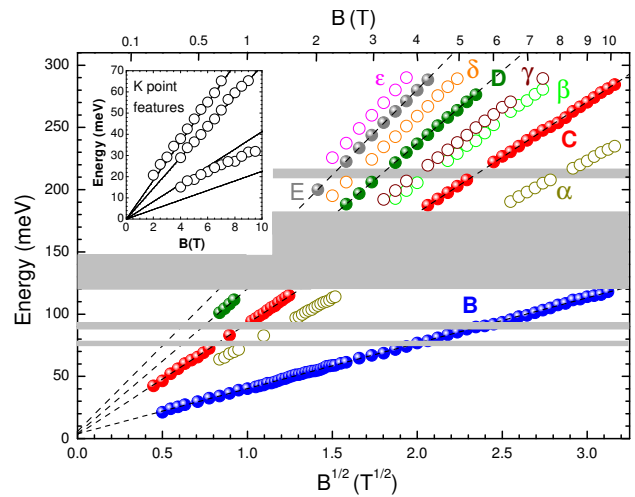}}
\caption{\label{LL} (color online) Positions of absorption lines related to the
$H$ point as a function of $\sqrt{B}$. The dashed lines represent a least squares
fit. Inset: $K$ point related transmission minima showing a nearly linear with
$B$ dependence, as compared to the most prominent transitions (solid lines with slopes
2.3, 4.1, 7.3 and 9.1 meV/T) anticipated after Ref.~\cite{LiPRB06}. The lowest
observed minimum is broad and can be composed of two lines.}
\end{figure}

The FIR experiments have been performed using the experimental
setup described in Ref.~\cite{SadowskiPRL06}. To measure the
transmittance of the sample, the radiation of globar, delivered
via light-pipe optics to the sample and detected by a Si
bolometer, placed directly below the sample and cooled down to a
temperature of 2~K, was analyzed by a Fourier transform
spectrometer. All measurements were performed in the Faraday
configuration with the magnetic field applied along the $c$-axis
of the sample. All the spectra were taken with non-polarized light
in the spectral range of 10-300~meV, limited further by several
regions of low tape transmissivity (see gray areas in
Figs.~\ref{SPKT} and \ref{LL}). The transmission spectra were
normalized by the transmission of the tape and by the zero-field
transmission, thus correcting for magnetic field induced
variations in the response of the bolometer.

Typical spectra, characteristic of a number of samples prepared in
the same way, are depicted in Fig.~\ref{SPKT}. Several absorption
lines showing a $\sqrt{B}$-dependence of their energies are observed
in the spectra, as seen in Fig.~\ref{LL}. We relate these
transitions to the $H$ point of graphite. The dominant lines in the
spectra are denoted by capital letters. The energies of the
subsequent lines B to E scale as
$1:(\sqrt{2}+1):(\sqrt{3}+\sqrt{2}):(2+\sqrt{3})$. The integral
intensities of these lines increase with the magnetic field. The
widths of the absorption lines also increase, see e.g. the C line in
Fig.~\ref{SPKT}, whose width increases from $\approx 10$~meV at
$B=1.3$~T to more than $30$~meV at $B=9.0$~T. Weaker intensity
features, which form a second series of absorption lines, are
denoted by Greek letters. In addition, features which shift linearly
with $B$, originating from the $K$ point of graphite, are also
observed in our spectra, see Fig.~\ref{SPKT}c. These features, which
have already been reported in magneto-reflectivity
data~\cite{LiPRB06}, are not further discussed in this work. The
weak intensity modulation of spectra (visible in Fig.~\ref{SPKT}a,c
above $B\approx 0.5$~T), which does not shift with changing magnetic
field, is attributed to interference effects in the sample.

\begin{figure}
\scalebox{0.72}{\includegraphics*[131pt,134pt][462pt,514pt]{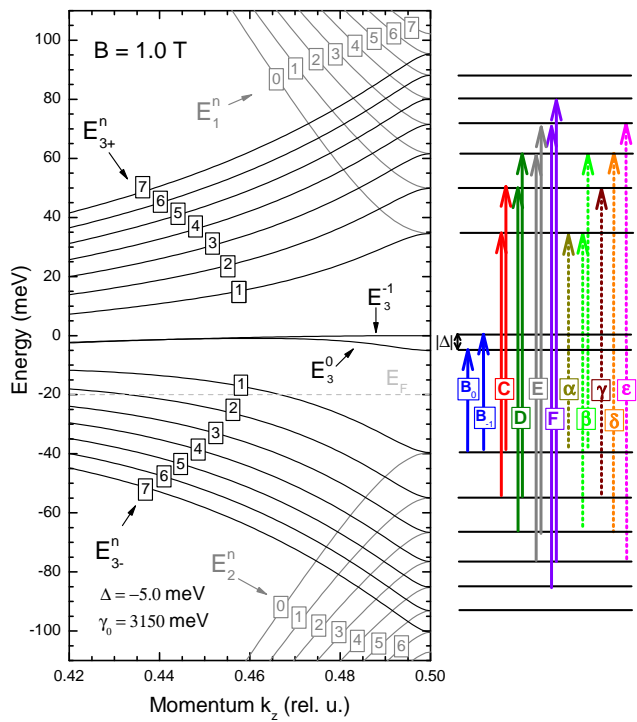}}
\caption{\label{Theory} (color online) LLs in graphite from $n=-1$
to 7 as a function of momentum $k_z$ in the vicinity of the $H$
point ($k_z=0.5$) at $B=1.0$~T calculated using the SWM model
neglecting trigonal warping ($\gamma_3=0$). The arrows on the
right-hand side show transitions considered in our interpretation.
Except for $\gamma_0$ and $\Delta$, the TB parameters used in the
calculation were taken from Ref.~\cite{MikitikPRB06}.}
\end{figure}

To interpret $\sqrt{B}$-dependent features in our FIR spectra,
we sketch the simplified SWM model
of Landau levels (LLs) in the vicinity of the $H$ point
\cite{McClurePR60,ToyPRB76}. Starting with four $\pi$-bands,
$E_1,E_2$ and the doubly degenerate $E_3$, we obtain four LLs
for each index $n\geq 1$ at a finite magnetic field $B$. In
addition, we obtain three levels and one level for $n=0$ and
$n=-1$, respectively. In the following, we use the notation
$E_1^n, E_2^n$ for $n\geq 0$, $E_{3+}^{n},E_{3-}^{n}$ for $n\geq
1$, $E_3^0$ and $E_3^{-1}$ to emphasize the band profile of the
individual LLs. To distinguish among levels in Fig.~\ref{Theory},
gray color is used for LLs $E_1^n$ and $E_2^n$. The assumption of
zero trigonal warping ($\gamma_3=0$)~\cite{McClurePR60,ToyPRB76}
simplifies the eigenvalue problem to the diagonalization of
$4\times4$ matrices. At $k_z=0.5$, we obtain the analytical
solution $E_3^{0}=\Delta$, $E_3^{-1}=0$ and for $n \geq 1$:
\begin{equation}\label{EnergyLL}
E_{3\pm}^{n}=E_{1,2}^{n-1}=\frac{\Delta}{2}\pm\sqrt{\frac{\Delta^2}{4}+\xi B
n},
\end{equation}
where ``$+$" and  ``$-$" correspond to $E_{1}^{n-1}$ and
$E_{2}^{n-1}$, respectively, and $\xi$ is related to $\gamma_0$
via the expression $\xi=3\gamma_0^2 e a_0^2/(2\hbar)$, with
$a_0=0.246$~nm~\cite{ChungJMS02}.

The selection rules for dipole-allowed interband transitions at $k_z=0.5$
predict the absorption lines at energies~\cite{ToyPRB76} ($n\geq 1$):
\begin{equation}\label{Interband}
\hbar\omega_n=\sqrt{\frac{\Delta^2}{4}+\xi Bn}+\sqrt{\frac{\Delta^2}{4}+\xi
B(n+1)},
\end{equation}
which correspond to transitions $E_{3-}^{n}$$\rightarrow$$
E_{3+}^{n+1}$ together with  $E_{2}^{n-1}$$\rightarrow$$E_{1}^{n}$
and, due to the expected electron--hole symmetry, also to
transitions $E_{3-}^{n+1}$$\rightarrow$$ E_{3+}^{n}$ and
$E_{2}^{n}$$\rightarrow$$E_{1}^{n-1}$. In addition to interband
transitions, two dipole-allowed intraband transitions
$E_2^0$$\rightarrow$$E_3^{-1}$ $(\hbar\omega_{-1})$ and $E_{3-}^1
$$\rightarrow$$ E_3^{0}$ $(\hbar\omega_{0})$, split in energy by $\Delta$,
\begin{equation}\label{Intraband}
\hbar\omega_{0} = \hbar\omega_{-1}+\Delta  =
\frac{\Delta}{2}+\sqrt{\frac{\Delta^2}{4}+\xi B},
\end{equation}
were experimentally found and discussed in Ref.~\cite{ToyPRB76}.

A number of dipole-allowed transitions expected at $k_z=0.5$ are
shown on the right-hand side of Fig.~\ref{Theory} by vertical solid
arrows. Following the general assumption that the Fermi energy at
the $H$ point is negative, in Fig.~\ref{Theory} we assign the
calculated lines to the experimentally observed lines, using the
same notation by capital letters B,C,D,E and F as in Fig.~\ref{SPKT}. It is important to note that
a non-zero value for the parameter $\Delta$ implies the splitting
of the strongest B transition into two B$_{0}$ and B$_{-1}$
components. However, this splitting is not resolved in the
experiment. The upper limit of the pseudogap, $|\Delta| < 10$~meV,
estimated from the measured width of the B line is in a good
agreement with Ref.~\cite{ToyPRB76}, but in an apparent
contradiction to Ref.~\cite{LiPRB06} and recent theoretical
calculations in Ref.~\cite{GruneisPRL08}. In Fig.~\ref{Theory}, we
have used $\Delta =-5.0$~meV. The clear observation of the B line,
which accurately scales as $\sqrt{B}$ down to $B\approx 0.3$~T,
allows us to estimate a Fermi energy, $E_F\approx -20$~meV, in
accordance with generally accepted values, see
e.g~\cite{MikitikPRB06}, but somewhat smaller than the ARPES
result ($E_F\approx-50$~meV)~\cite{ZhouNatPhys06}.

Another important feature of our data is the clearly asymmetric
shape of the B line as well as the significant line broadening
with increasing magnetic field. We presume that both these effects
indicate some contribution of the transitions at $k_z\neq0.5$ to
the observed transmission spectra. At low magnetic field, the
$k_z$-dispersion in the vicinity of the $H$ point is steeper for
$E_{2}^0$ than for $E_{3-}^1$ and this qualitatively explains the
low field asymmetry. The dispersions become more symmetric with
increasing $B$, as does the line shape of the B line.
Nevertheless, the dispersions of all levels are steeper at higher
magnetic fields, which qualitatively accounts for the broadening
of the absorption lines. However, the agreement remains only
qualitative. An additional mechanism for the line broadening is
electron scattering probability, which increases linearly when
moving away from the Dirac point~\cite{ZhengPRB02}. Summing over
the transitions for different momenta $k_z$, in terms of a simple
calculation of the joint density of states, does not help to
reproduce the spectral shapes of the observed lines. Some
additional $k_z$ dependent selection rules, enhancing the
oscillator strength in the close vicinity of $k_z=0.5$ point need
to be introduced.

The relatively small value of the pseudogap $\Delta$ is consistent
with the very good linearity of data with $\sqrt{B}$ (see
Fig.~\ref{LL}). The limit of $\Delta\rightarrow 0$ allows us to
rewrite Eq.~\eqref{EnergyLL} as $E_{3\pm}^{n}=E_{1,2}^{n-1} =
\pm\sqrt{\xi Bn}$, which has the form of LLs in a 2D system of
massless Dirac fermions. The energies of the dipole-allowed
transitions~\eqref{Interband},\eqref{Intraband} can then be for
$n\geq0$ written as $\hbar\omega_n=\sqrt{\xi
B}(\sqrt{n}+\sqrt{n+1})$  and a clear correspondence to the
optical transitions in graphene is
established~\cite{SadowskiPRL06,GusyninPRL07}. The Fermi velocity
can then be expressed in the form $\tilde{c}=\sqrt{\xi/(2e\hbar)}$
and from our data evaluated to be
$\tilde{c}=(1.02\pm0.02)\times10^6$~m/s, which is about 10\%
higher than in~\cite{ZhouNatPhys06} but in a good agreement with
the most recent data~\cite{LiNatPhys07,GruneisPRL08}. We stress
that the entire fan chart of the observed inter LL transitions
reported here can be described with a single parameter (Fermi
velocity). This is a surprising observation, as Kohn's theorem is
not expected to hold in a system with a strongly non parabolic one
particle dispersion, so that electron-electron interaction may
differently alter excitations between different pairs of LLs. Note
that the linear fits in Fig.~\ref{LL} extrapolate to an onset
$\approx3$~meV instead of zero, which possibly suggests, apart from
the anticrossing splitting $\Delta$, an additional
mutual shift of the electron and hole Dirac cones, however, this
effect lies on the edge of the experimental accuracy.

We turn now our attention to the additional transitions, denoted
in~Figs.~\ref{SPKT} and \ref{LL} by Greek letters, which are in
general of a lower intensity. They exhibit a clear
$\sqrt{B}$-dependence and have no counterpart in the transmission
spectra of graphene~\cite{SadowskiPRL06,JiangPRL07}. To interpret
these lines we have to go beyond the simplified model used in
~\cite{McClurePR60,ToyPRB76} and consider a more complete
analysis~\cite{DresselhausPR65}. The selection rules
$n$$\rightarrow$$n\pm 1$ ($n\geq1$) is predicted not only for the main
transitions $E_{3-}^{n}$$\rightarrow$$ E_{3+}^{n\pm1}$ and
$E_{2}^{n}$$\rightarrow$$ E_{1}^{n\pm1}$, but also for a second
series of lines, presumably weaker in intensity,
$E_{3-}^{n}$$\rightarrow$$ E_{1}^{n\pm1}$ and
$E_{2}^{n}$$\rightarrow$$ E_{3+}^{n\pm1}$. We can then relate the
absorption lines $\alpha,\gamma,\delta$ and $\epsilon$ to
transitions which are symmetric with respect to the Dirac point,
as depicted in Fig.~\ref{Theory} by the vertical dotted arrows. In
agreement with expectations, the energies of these lines scale as
$1:\sqrt{2}:\sqrt{3}:2$ and the same Fermi velocity is derived
($\tilde{c}=(1.02\pm0.02)\times10^6$~m/s) as for the main
transitions.

Another series of weak absorption lines, predicted
in~\cite{DresselhausPR65}, should satisfy the selection rules
$n$$\rightarrow$$n\pm 2$. Their oscillator strength is directly
connected to the parameter $\gamma_3$. The $\beta$ line can then
be assigned to transitions
$E^{3(1)}_{3-}$$\rightarrow$$E^{1(3)}_{3+}$ and/or
$E^{2(0)}_{2}$$\rightarrow$$E^{0(2)}_{1}$, representing thus  a
direct indication of the trigonal warping in graphite,
demonstrated experimentally e.g.
in~\cite{ZhouNatPhys06,GruneisPRL08}.

Having interpreted all observed absorption lines, we can draw the
following conclusions. The dominant transitions in the spectra,
which have their counterpart in the spectra of
graphene~\cite{SadowskiPRL06,JiangPRL07}, justify that holes in
HOPG can, with a reasonable accuracy, be described as a purely 2D
gas of Dirac fermions. This justifies the model adopted for
interpreting STS experiments~\cite{LiNatPhys07} and
quantum oscillations in
graphite~\cite{LukyanchukPRL04,LukyanchukPRL06}. Nevertheless,
the latter analysis is not consistent with our data, when estimating the
hole density. Whereas in our experiment, the hole filling factor
$\nu=6$ is achieved at $B\approx 0.3$~T, when the B line clearly
appears in the spectrum, the same value of $\nu$ is found at
magnetic fields above 1~T in the magnetotransport
data~\cite{LukyanchukPRL04,LukyanchukPRL06}.

On the other hand, the series of additional absorption lines of
weaker intensity, which are dipole-forbidden in the strictly 2D
gas of Dirac particles, show that in HOPG we are dealing with a
strongly anisotropic but nevertheless 3D system and the model of a
purely 2D gas is not valid in general. The 3D nature of graphite
is also shown by the pronounced low field asymmetry of the B line
and possibly by the observed increase in linewidth with increasing
$B$.

In summary, FIR magneto-transmission measurements have been used
to probe the Dirac holes at the $H$ point in a thin graphite layer
prepared by the exfoliation of HOPG. We find a relatively small
value for the pseudogap $|\Delta|<10$~meV, consistent with the
observation of Dirac fermions at the $H$ point of graphite, with a
Fermi velocity of $\tilde{c}=(1.02\pm0.02)\times10^6$~m/s. The
main absorption lines can be understood using a model which
assumes a strictly 2D gas of Dirac fermions, giving evidence for
the strong anisotropy of HOPG. Nevertheless, the presence of
additional weaker transitions, dipole-forbidden in a purely 2D
system of Dirac fermions, can only be understood by considering
the 3D nature of graphite.

\begin{acknowledgments}
The present work was supported by the European Commission through
Grant No. RITA-CT-2003-505474, by contract ANR-06-NANO-019 and projects
MSM0021620834 and KAN400100652.
\end{acknowledgments}

\bibliography{Orlita_et_al-ref}
\end{document}